\documentclass[journal]{IEEEtran}
\usepackage{graphicx}
\usepackage{epsf}
\usepackage{mathtools}
\usepackage{amssymb}
\usepackage{amsmath, amsfonts}
\usepackage{latexsym}
\usepackage{multirow}
\usepackage{multicol}
\usepackage[table]{xcolor}
\usepackage{color}

\usepackage{tabularx}
\usepackage{lipsum}

\usepackage{algorithm2e}
\RestyleAlgo{ruled}
\SetKwComment{Comment}{// }{}

\usepackage{enumerate}
\usepackage{enumitem}
\usepackage{tikz}

\usepackage{soul}

\usepackage{url}

\usepackage{mathrsfs}

\usepackage{fixltx2e}

\usepackage[mathscr]{euscript}

\usepackage{commath}
\usepackage{MnSymbol}

\usepackage{subcaption}

\usepackage{mathtools} 
\DeclarePairedDelimiterX\setc[2]{[}{]}{\,#1 \;\delimsize\vert\; #2\,}
\DeclarePairedDelimiterX\parth[2]{(}{)}{\,#1 \;\delimsize\vert\; #2\,}

\PassOptionsToPackage{hyphens}{url}
\usepackage[unicode=true, bookmarks=true, bookmarksnumbered=true, bookmarksopen=true, bookmarksopenlevel=1, breaklinks=false, pdfborder={0 0 0}, pdfborderstyle={}, backref=false, colorlinks=false]{hyperref}

\usepackage{courier}

\usepackage{theorem}
{
\theorembodyfont{\rmfamily}

}

\usepackage[nospace]{cite}

\definecolor{orange}{RGB}{255,127,0}
\definecolor{blue}{RGB}{0,0,255}
\definecolor{red}{RGB}{255,0,0}
\definecolor{green}{RGB}{50,160,50}
\definecolor{grey}{RGB}{125,120,125}
\definecolor{purple}{RGB}{125,0,125}

\begin{document}
{
\title{{\fontsize{18}{2}\selectfont Optimizing EV Chargers Location via Integer Programming}}

\author
{
Seungmo Kim, \textit{Senior Member}, \textit{IEEE}, Yeonho Jeong, \textit{Member}, \textit{IEEE}, and Jae-Won Nam, \textit{Member}, \textit{IEEE}

\vspace{-0.3 in}

\thanks{S. Kim is with the Department of Electrical and Computer Engineering, Georgia Southern University in Statesboro, GA, USA. Y. Jeong is with the Department of Electrical and Computer Engineering, University of Rhode Island in Kingston, RI, USA. J.-W. Nam is with the Department of Electronic Engineering, Seoul National University of Science and Technology, Seoul, South Korea. The corresponding author is J.-W. Nam who can be reached at jaewon.nam@seoultech.ac.kr. This study was financially supported by Seoul National University of Science and Technology.}
}

\maketitle
\begin{abstract}
There is no question to the fact that electric vehicles (EVs) are the most viable solution to the climate change that the planet has long been combating. Along the same line, it is a salient subject to expand the availability of charging infrastructure, which quintessentially necessitates the optimization of the charger's locations. This paper proposes to formulate the optimal EV charger location problem into a facility location problem (FLP). As an effort to find an efficient method to solve the well-known nonpolynomial deterministic (NP)-hard problem, we present a comparative quantification among several representative solving techniques.
\end{abstract}

\begin{IEEEkeywords}
Electric vehicle (EV) charging; Facility location problem (FLP); Integer program; Optimization solvers
\end{IEEEkeywords}

\section{Introduction}\label{sec_intro}
\subsubsection{Background}
It is an easily agreeable subject that EVs are a promising solution to relieve environmental issues. In fact, EV sales in the United States (U.S.) have increased yearly. Charging infrastructure is critical to the continued growth of EV and its upstream industries. A lack of convenient and ubiquitous charging infrastructure is one of the key factors that impedes EV adoption \cite{mckinsey_20}.

The U.S. federal government moved swiftly to address this. An example is the EV Charging Action Plan \cite{biden_bill,raise} that provides \$7.5 billion to develop 500,000 public chargers nationwide by the Year 2030.

The private sector has also responded positively to the government's leading effort. As an example, Tesla announced commitment to open thousands of its ``Superchargers'' to EVs made by other manufacturers \cite{tesla_23}. Currently, Tesla provides 28,000 charging ports at Supercharger stations in the U.S., which have been accessible primarily to drivers of the company's own cars until now.

Nonetheless, the reality still looks quite far-fetched. The U.S. government aims at the average availability of EV charging station every 50 miles \cite{ev_mile} (compared to gas stations existent every 3.5 miles \cite{gas_mile}!), while low-income residents are more dependent on automobiles and must travel further to access jobs and essential services \cite{equity_low income_21}.

To this line, this paper lays out a \textit{theoretical framework of optimal EV chargers deployment}.

\subsubsection{Contributions}
We emphasize that our EV charger location optimization problem will very highly likely take a wide diversity of variables (including economic, societal, human behavioral, etc.) into formulation. As such, we set this paper to contribute to:
\begin{itemize}
\item \fontsize{10}{11}\selectfont Building a \textit{comprehensive mathematical framework} accommodating the particular complexity;
\item \vspace{0.04 in} \fontsize{10}{10}\selectfont Demonstrating our numerical computational framework for solving the facility location problem (FLP) representing the optimal location
\item \vspace{0.04 in} \fontsize{10}{10}\selectfont Laying out an extensive comparative study among the optimization \textit{solving techniques} as an effort to find the most efficient solver
\end{itemize}

\section{Related Work}\label{sec_related}
It is known to be a challenging task to create accurate maps of utility infrastructure are important for planning and operations, managing risk, and rapidly assessing damages after a storm \cite{poles_mdpi18_NaM89}. However, the lack of exact locations of electric facilities is not uncommon \cite{poles_mdpi18_NgJ18}. For example, after hurricane Maria struck Puerto Rico in 2017, the lack of accurate maps for buildings, bridges, and electric facilities was considered as a main factor slowing recovery efforts \cite{poles_mdpi18_Cit17}. Mapping utility poles is labor- and time-intense because the process is usually conducted using human interpretation of high spatial-resolution aerial imagery, ground-based field surveys, or unmanned aerial vehicles (UAVs)/helicopters \cite{poles_mdpi18_NgJ18,poles_mdpi18_CeB09}. The high degree of labor requirement makes mapping utility poles over large areas a daunting task \cite{poles_mdpi18}.

\subsubsection{On Problem Formulation} Despite their importance \cite{poles_mdpi18_NaM89}, identifying locations of utility poles over a large area is often laborious and time-intensive due to the need for human interpretation on various types of data \cite{poles_mdpi18_NgJ18,poles_mdpi18_Cit17,poles_mdpi18_CeB09}.

Meanwhile, latest literature introduced a body of prior work that formulates the optimal deployment of EV chargers as a multi-objective optimization problem \cite{planning_qap_21}, considering various factors such as battery types \cite{qatar_20} and distances to nearby energy sources \cite{multiobjective20}.

We extended our investigation to the literature of facility location problem (FLP),  which is analogous to our problem in the sense of having to determine optimal deployment plans including locations and expansion patterns \cite{splp_83,wiki_CoM14}. It took our attention that the FLP is formulated as the mixed integer programming \cite{distribution planning_optimal path_21,distribution planning_optimization_90}, wherein the constraints are reduced to continuous-variable linear equations. The literature goes on to generalization of the formulation to the mixed integer linear programming \cite{distribution planning_reliability_11,ip_linearization_05} and the mixed-integer nonlinear programming (MINLP) \cite{distribution planning_multi-stage_18}.

\subsubsection{On Solving Techniques} The key challenge in this research is an exhaustive enumeration that would quickly become computationally hopelessly expensive for the MINLP \cite{ip_difficult,qap_survey_20,qap_survey_20_SaG76}. As such, solving a MINLP in polynomial time by using \textit{exact} algorithms (e.g., cutting plane, branch and bound, etc.) can only be considered for small instances. For large instances, it is difficult to enumerate all the solutions due to the number of permutations that easily explodes to $n!$ \cite{qap_survey_18}, leaving fewer options such as the black-box solver \cite{qap_survey_20_Fis12,qap_survey_20_Zha13}.

Thus, for solving such large-instance NP-hard problems, \textit{heuristic} approaches attract broader interests \cite{qap_survey_18}. Considering our MINLP problem that is uniquely characterized by a large number of variables, we find it particularly suitable to adopt the \textit{metaheuristic}, which  are known particularly efficient in solving combinatorial optimizations (which this research seeks to solve) by searching over a large set of feasible solutions with less computational effort, especially with incomplete or imperfect information or limited computation capacity \cite{metaheuristics_03}. Metaheuristics sample a subset of solutions which is otherwise too large to be completely enumerated or otherwise explored \cite{metaheuristics_09}. We performed thorough investigation of the literature on metaheuristics, including the tabu search (TS) \cite{qap_survey_20_Tai91,qap_survey_20_Dre05b}, random swaps \cite{qap_survey_20_Misevicius12}, genetic algorithm \cite{qap_survey_20_Dre03,qap_survey_20_Dre05a}, simulated annealing (SA) \cite{qap_survey_20_Chi98,qap_survey_20_Mis03,qap_survey_20_Mis04}, ant colony \cite{qap_survey_20_Gam99,qap_survey_20_TaR01}, and memetic algorithms \cite{qap_survey_20_Ben15,qap_survey_20_Har15}.

\section{Problem Formulation}\label{sec_formulation}
Here is how our problem is uniquely defined. The optimal locations of EV chargers will be found via solving a \textit{FLP} \cite{splp_83}. We modify the traditional FLP such that it can further accommodate a wide diversity of factors (including economic, societal, human behavioral, etc.) depending on the context, to which the problem is applied.

\subsection{Spatial Setup}\label{sec_formulation_spatial}
We characterize the distribution of EV chargers as a \textit{homogeneous Poisson point process (PPP)} over a finite two-dimensional space $\mathbb{R}^2$. As shall be detailed in Section \ref{sec_results_flp}, we deploy \textit{EV chargers} and \textit{demanding areas}, and find the connections from an EV charger to (a) demanding area(s). Predicated on the assumption of PPP, the locations of the chargers and demanding areas follow a \textit{uniform distribution} on both X and Y axes.

This spatial setup forms the foundation for the key optimization problems that this paper targets to solve: viz., FLP for the optimal location.

\subsection{Formulation to Capacitated FLP}\label{sec_formulation_flp}
Suppose there are $n$ facilities and $m$ customers. One wishes to choose (i) which of the $n$ facilities to open, and (ii) which of the open facilities to use to supply the $m$ customers, in order to satisfy some fixed demand at minimum cost. We propose to modify the cannonical form of \textit{capacitated FLP} \cite{splp_83} into a MINLP, which is given by
\begin{align}\label{eq_flp}
&\text{min } \overbrace{\sum_{i=1}^{n}\sum_{j=1}^{m} v_{ij}d_{ij}y_{ij} + \sum_{i=1}^{n} s_{i}x_{i}}^{\text{\scriptsize Classical capacitated FLP \cite{splp_83}}}\nonumber\\
&\hspace{0.8 in}+ \underbrace{\sum_{i=1}^{m}\sum_{j=1}^{n} \mathsf{E}_{ij}^{-1} + \sum_{i=1}^{m}\sum_{j=1}^{n} \mathsf{\Theta}_{ij}\left(\theta_{1}, \cdots, \theta_{N}\right) + \cdots}_{\text{\scriptsize Via this work}}
\end{align}
\begin{align}
&\left.
    \begin{array}{ll}
        \text{s.t. } \sum_{i=1}^{n} y_{ij} \le 1\\\nonumber
\hspace{0.25 in} y_{ij} \ge 0, \hspace{0.1 in} \forall i,j = 1, \cdots, n\\\nonumber
\hspace{0.25 in} x_{i} \in \left\{0,1\right\}, \hspace{0.1 in} \forall i = 1, \cdots, n\nonumber
    \end{array}
\hspace{0.01 in} \right \}\text{\scriptsize Classical capacitated FLP \cite{splp_83}}\nonumber
\end{align}
\begin{align}
&\left.
    \begin{array}{ll}
\hspace{0.43 in} \mathsf{C}_{\text{min}}x_{i} \le \sum_{j=1}^{m} d_{j}y_{ij} \le \mathsf{C}_{i}x_{i}, \hspace{0.1 in} \forall i = 1, \cdots, n\\\nonumber
\hspace{0.43 in}0 \le \mathsf{E}_{ij}^{-1} \le 1, \hspace{0.1 in} \forall i,j = 1, \cdots, n\\\nonumber
\hspace{0.43 in} \text{Any additional constraints on } \theta_{1}, \cdots, \theta_{N}\nonumber
    \end{array}
\hspace{0.01 in} \right \}\text{\scriptsize Via this work}
\end{align}

\vspace{0.05 in}
We identify the parameters of our modified FLP as below:
\begin{itemize}[parsep=0pt,itemsep=0pt]
\item \vspace{0.03 in} \fontsize{10}{10}\selectfont $i$ and $j$ are indexes for an EV charging facility and a demanding area (or, equivalently, a customer), respectively
\item \vspace{0.04 in} \fontsize{10}{10}\selectfont $v_{ij}$ gives the \textit{variable cost} to get the electricity supplied to serve customer $j$
\item \vspace{0.04 in} $d_{j}$ gauges the \textit{demand} from customer $j$
\item \vspace{0.04 in} \fontsize{10}{10}\selectfont $y_{ij}$ quantifies the fraction of the demand made by customer $j$ and fulfilled by facility $i$
\item \vspace{0.04 in} $x_{i}$ indicates whether facility $i$ \textit{opens or not}
\item \vspace{0.04 in} \fontsize{10}{10}\selectfont $s_{i}$ denotes the \textit{sunken cost} (also known as ``fixed'' cost) of opening a charging facility $i$
\item \vspace{0.04 in} \fontsize{10}{10}\selectfont $\mathsf{E}_{i,j}$ defines the \textit{equity} achieved at customer $j$ via service from facility $i$
\item \vspace{0.04 in} \fontsize{10}{10}\selectfont $\mathsf{C}_{i}$ and $\mathsf{C}_{\text{min}}$ indicate the \textit{capacity} of facility $i$ and the required minimum capacity of any facility, respectively, both in the unit of kWh.
\end{itemize}

\vspace{0.05 in}
To elaborate on a few key variables, $\mathsf{E}_{ij}$ measures the equity by using the \textit{Gini coefficient}, which ranges from 0 (i.e., complete equality) to 1 (i.e., complete inequality). As a measurement for the inequality of wealth or income \cite{gini13}, the Gini coefficient has also been used to measure how evenly the resource is allocated to the participants in a network \cite{gini}. We propose to use the coefficient as a gauge of \textit{how many of the demands around a facility} are addressed.



As another means to pursue the equity, we propose a constraint with the minimum capacity for any charger, $\mathsf{C}_{\text{min}}$. In fact, the State of Georgia has also adopted this idea in their EV charger deployment plan \cite{ga_corridor}. Moreover, it has been found that integrating a multitude of chargers at a single facility can contribute to lowering the sunken cost $s_{i}$ \cite{charging_cost_19}.

By $\mathsf{\Theta}_{ij}$, we leave some room for possible \textit{addition of new variables} as the formulation evolves to reflect the reality more accurately. As a MINLP, the formulation given in Eq. (\ref{eq_flp}) can accommodate $\mathsf{\Theta}_{ij}$ either in the form of linear or nonlinear. Also, each variable $\theta_{1}, \cdots, \theta_{N}$ can either be discrete or continuous.

\subsection{Unique Challenges}\label{sec_formulation_challenges}
We are aware that the MINLP is a well-studied area over the last few decades. In particular, the size and complexity of IP problems being successfully solved has ever increased, mostly thanks to the continued development of relevant algorithms---e.g., branch and bound \cite{ip_practical_solution_74}.

Nonetheless, we consider this research novel and significant, owing to the extreme complexity of the target problem. The complexity is mainly attributed to the following key reasons \cite{ip_challenging_81}:
\begin{itemize}
\item \vspace{-0.02 in} \fontsize{10}{10}\selectfont \textsf{\textbf{C1:}} \textit{Large search spaces} for domain and other variables
\item \fontsize{10}{10}\selectfont \textsf{\textbf{C2:}} \textit{Inexistence of polynomial-time} numerical solving techniques
\end{itemize}

In regard to challenge \textsf{\textbf{C1}}, the focus of this research is to deal with a \textit{large number of variables}, which will be unavoidable to precisely quantify the equity $\mathsf{E}_{ij}$ reflecting all the demographic, geographic, economic factors. The challenge here is the \textit{dissimilarity} among the different data. As a remedy, we build on the literature of \textit{coupled matrix and tensor factorization (CMTF)}, which jointly factorizes multiple datasets in the form of higher-order tensors and matrices by extracting a common latent structure from the shared mode \cite{cmtf_elsevier15,cmtf_ieee14,cmtf_plos19,cmtf_ieee21}.

Another focal point of this work is \textit{addition of constraints} as a means to (i) cut off infeasible solutions  \cite{ip_linearization_05_SaG91,ip_linearization_05_Lin02,ip_linearization_05_BuF02} and/or (ii) linearization \cite{ip_linearization_mdpi22,ip_linearization_05_Glo75,ip_linearization_05_Flo95}. However, we will be especially sensitive in adding cuts to Eq. (\ref{eq_flp})---which is already NP-hard. The reason is, while it removes integer infeasibilities, it can incur more constraints in each node of a branch-and-bound process, which can cause a higher delay in solving \cite{ip_difficult}.

As a response to challenge \textsf{\textbf{C2}}, this we focus on keeping our optimization framework in a \textit{flexible} form, which is quintessential to accommodate more variables (some of which are even \textit{unknown}!) as framework evolves to reflect the real-world characteristics of our problem. In reality, many factors gauging the societal/economic/demographic equities can dynamically \textit{change} both spatially and temporally as the society evolves \cite{equity_mit21,healthequityUS_19,uncertainty_19,uncertainty_19_Ali17b}. This crucial necessity of flexibility significantly compounds the complexity to our formulation, to which we propose to understand the feasibility of countermeasure techniques including iterative methods \cite{hard_co_20}, hierarchically structured space \cite{dependency_tree_97}, sensitivity analysis \cite{uncertainty_19_Hyd05}, fuzzy decision analysis \cite{uncertainty_19_Zar08, uncertainty_19_Afs11, uncertainty_19_Est16}, and Monte-Carlo selection methods \cite{uncertainty_19_Mad11,uncertainty_19_Mad14}.

Particularly in our problem, the factors forming the objective and constraints can dramatically change both spatially and temporally. That is, many factors gauging the societal/economic/demographic equities \cite{qatar_20,equity_mit21} are subject to form ``temporality'' and hence change over time as the society evolves \cite{healthequityUS_19}. This makes a strong case that the optimization problem and the solving method must be formulated into flexible forms so they can accommodate any addition/removal/change of parameters. It is another critical factor that makes our formulation more complicated.

\section{Solving Techniques Development}\label{sec_solving}

\subsection{Unique Challenges and Proposed Approaches}\label{sec_solving_proposed}
Recall that the FLP formulated as (\ref{eq_flp}) is a NP-hard problem \cite{wiki_FoP81,wiki_MeT82}. This entails critical challenges in \textit{solving} the problem.

We have already identified the metaherustic as the primary method to implement our problem (\ref{eq_flp}), which would be the most suitable option considering the unique challenges of our problem \textbf{\textsf{C1}} and \textbf{\textsf{C2}}. {\color{blue} In our problem}, it is particularly important to find an approximate global optimum than to find a precise local optimum in a fixed amount of time, which makes a compelling case where the SA is preferable than exact algorithms (such as gradient descent, branch and bound, etc.) \cite{sa_KiG83}.

Furthermore, algorithms with an aim to solve large-size instances of such combinatorial optimization problems apply \textit{parallelism} for expedition of both exact methods (e.g., the branch-and-bound algorithms) \cite{qap_survey_20_Par89,qap_survey_20_ClP97} as well as heuristics \cite{ip_difficult}. A representative example for the latter is parallelization of the objective function evaluation during a tabu search \cite{qap_survey_20_Luo11}, for which graphical processing units (GPUs) \cite{qap_survey_20_Tsu09,qap_survey_20_Pov18,qap_survey_20_Son18} as well as the Compute Unified Device Architecture (CUDA) platform \cite{cuda_nvidia,qap_survey_20_Cza13} have been used. As a means to efficient memory management \cite{qap_survey_20_Zhu10}, this research also seek the feasibility of \textit{cooperative heuristics} with a particular aim of improving the solution's speed and accuracy. The parallel instances can be executed via a global memory \cite{qap_survey_20_TaB06,qap_survey_20_Dok15,qap_survey_20_Tos15,qap_survey_20_MuD16,qap_survey_20_JaR05,qap_survey_20_JaR09a,qap_survey_20_AkD17,qap_survey_20_AbD19,qap_survey_20_DoC16}, or via distributed-memory systems \cite{qap_survey_20_ClP97}. It is noteworthy that the latter is efficient when the algorithms are independent among the parallel instances and thus no exchange of information across the memories is critical.

Many variants of the integer program (IP) are acknowledged to be NP-hard \cite{co_GhG04}. With this noted, this paper is devoted to developing \textit{computational tools} for solving the proposed FLP problem, a well-acknowledged NP-hard IP problem \cite{tsp_ip_74}.

\subsection{Comparison among Solving Techniques}\label{sec_solving_comparison}
For a dedicated purpose of comparing a variety of numerical optimization solving techniques, we propose to modify Eq. (\ref{eq_flp}) into an abstract form of the \textit{integer NLP}. Especially, we notice that the complexity is particularly induced from the nonlinearity. Thus, it is of critical importance to come up with a test function that suits to test the nonlinearity accurately.

We identify the \textit{Rastrigin function} \cite{ras_74} that has long been known as a representative example, through which the performance comparison among the multitude of solving techniques can be clarified \cite{ras_91}. Considering the unique nature of containing a large number of variables defining the objective and constraints, we expand the Rastrigin function such that $n$ is a sufficiently large number:
\begin{align}
f_{\text{ras}}\left(\mathbf{x}\right) = An + \sum_{i=1}^{n} \left[x_{i}^2 - A\cos\left(2\pi x_{i}\right)\right]
\label{eq_ras}
\end{align}
where $A=10$ and $x_{i} \in \left[-5.12, 5.12\right]$.

\subsection{Alternative Techniques}\label{sec_solving_alternative}
We shed light on alternative approaches, considering that the combinatorial optimization is notorious to be highly complex and thus one may end up having to find an ``approximate'' solution to the global optimal. Yet, we reiterate that SA, the method that we propose to adopt to solve the proposed optimization problem, is acknowledged for its ability to solve very complicated optimization problems even when exact methods fail \cite{sa_DuS02}. Thus, it can still suffice what this research is looking for, in a practical sense.

Even despite this safe selection of method, in the event the optimal solution too widely varies, we plan to adopt statistical techniques that will help uphold the reliability of the proposed SA mechanism. An example technique is the principal component analysis (PCA) \cite{pca87}, which seeks to identify a certain set of factors that particularly dominate the solution of the QAP.

Moreover, it was found that machine learning techniques can be applied to efficiently solving combinatorial optimizations. Examples include reinforcement learning \cite{neural_arxiv16,survey21}, to which the authors have already been piling a significant body of contributions \cite{arxiv2004,arxiv2101,tiv22,vtc20f,vtc21s,iceic22,nasim22}.

Parallel computing has also been attracting a considerable amount of research interest, thanks to its ability to distribute an extremely complex (and thus hopelessly challenging to solve) optimization problem into smaller instances and solve them instead. As such, it can be deemed an efficient strategy to utilize a parallel computing cluster. As an option for a further scale-up to a larger pool of servers, we suggest to use the MATLAB Parallel Server \cite{matlab_parallel}.

\begin{figure}[t]
\centering
\includegraphics[width = \linewidth]{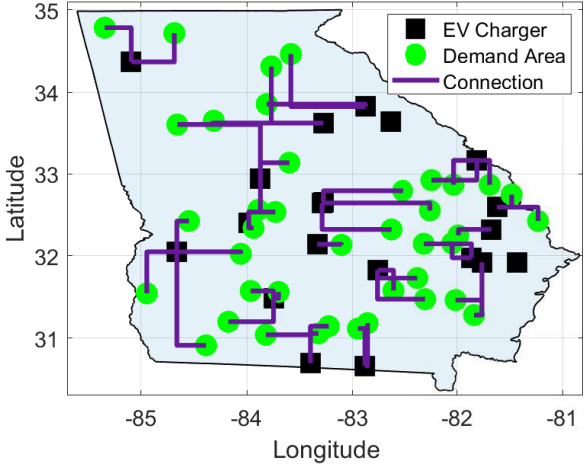}
\caption{Example mapping of 20 EV chargers and 40 demand areas on the map of the State of Georgia}
\label{fig_flp_map}
\end{figure}

\begin{figure}[t]
     \centering
     \begin{subfigure}[b]{\linewidth}
         \centering
         \includegraphics[width=\textwidth]{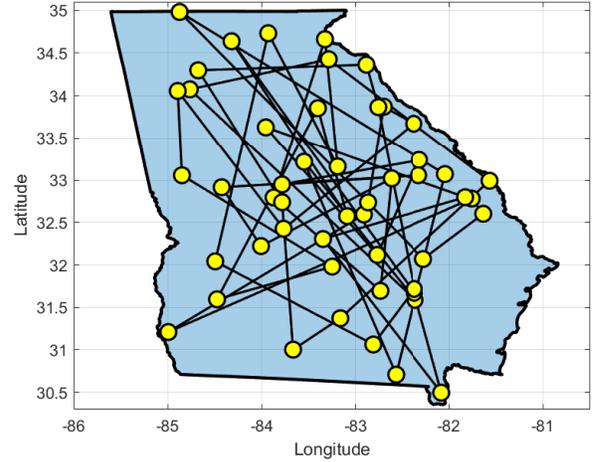}
         \caption{After 1 iteration}
         \label{fig_tsp_sa_iter1}
     \end{subfigure}
     \hfill
     \begin{subfigure}[b]{\linewidth}
         \centering
         \includegraphics[width=\textwidth]{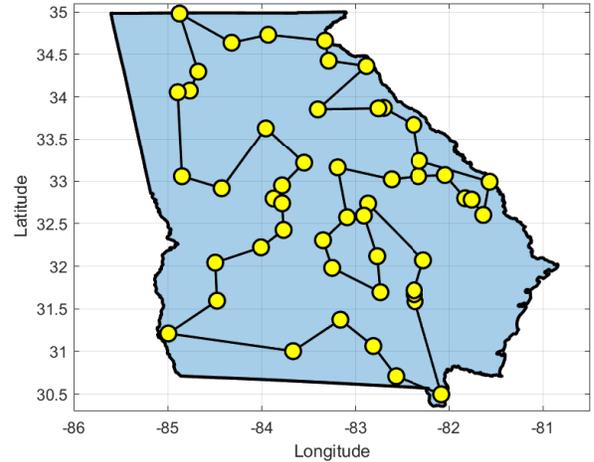}
         \caption{After 300 iterations}
         \label{fig_tsp_sa_iter300}
     \end{subfigure}
\caption{Solving a TSP by using SA (The agent (i.e., salesperson) has to visit $n=50$ points created on a map of the State of Georgia. The map is with shorter connections as iterations progress through the course of SA.)}
\label{fig_tsp_sa}
\end{figure}

\begin{figure}[t]
\centering
\includegraphics[width = \linewidth]{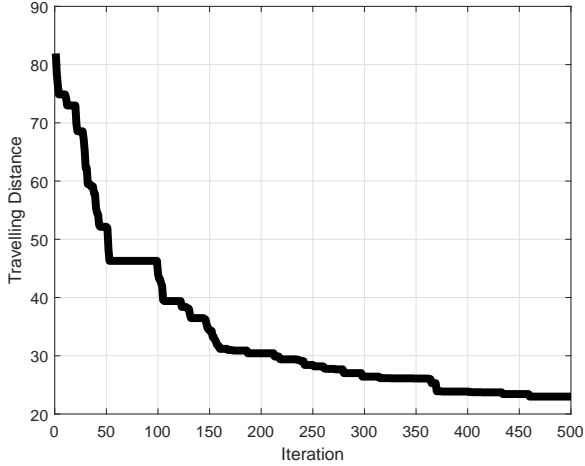}
\caption{Convergence of total travelling distance over the route}
\label{fig_tsp_distance}
\end{figure}

\begin{table*}[t]
\centering
\caption{Comparison among several representative optimization solvers}
\begin{tabular}{|c || c c c c | c | c|}
\hline
\rowcolor{gray!20} \textbf{Solver} & \textbf{$x_{1}$} & \textbf{$x_{2}$} & \textbf{$\cdots$} & \textbf{$x_{10}$} & \textbf{Objective Value} & \textbf{Number of Iterations} \\ [0.5ex]\hline\hline
Integer Linear Programming & 4.4409e-16 & 4.4409e-16 & $\cdots$ & 4.4409e-16 & 0 & 0\\\hline  
Pattern Search & 0 & 0 & $\cdots$ & 0 & 0 & 204\\\hline
Genetic Algorithm & -0.062657 & 0.042974 & $\cdots$ & -0.041941 & 1.4801 & 3907\\\hline
Particle Swarm & -7.2517e-07 & 2.5503e-08 & $\cdots$ & 1.7757e-06 & 7.3e-10 & 4320\\\hline
Simulated Annealing & 6.4039e-05 & -1.99 & $\cdots$ & 0.00018799 & 3.9798 & 3008\\\hline
Surrogate Optimization & 0.99678 & 1.9937 & $\cdots$ & 1.9832 & 8.9671 & 200\\\hline
\end{tabular}
\label{table_compare}
\end{table*}

\section{Numerical Results}\label{sec_results}
Now, we present the experimental results. We undertook extensive computer simulations on the following three fronts.

\subsubsection{Solving FLP}\label{sec_results_flp}
The first is to solve the FLP \cite{splp_83} by using an ILP solver. Fig. \ref{fig_flp_map} depicts an example mapping between 20 chargers and 40 demand areas, which are generated at random locations following a homogeneous PPP. The purple lines indicate which charger serves which demand areas. Note that the focus of this particular simulation is put showing how the problem is solved, rather than how to generate optimal facility locations. We used the MATLAB's \texttt{intlinprog} solver, which finds minimum of a constrained integer linear multivariate optimization.

\subsubsection{Solvers Comparison}
As an effort to lay out a broader perspective on solving our problem, we make a comparison among a variety of numerical optimization methods. It is noteworthy that we refer to the Rastrigin function that has earlier been shown in Eq. (\ref{eq_ras}) with $n=10$ as an effort to reflect the ``many-variable'' nature of our problem. Table \ref{table_compare} lays out the comparison. Note that column ``Objective Value'' shows the optimal value of the objective function, and columns $x_{1}$ through $x_{10}$ indicate the values of $x_{i}$'s in Eq. (\ref{eq_ras}) yielding the optimum.

\subsubsection{Route Minimization into Account}\label{sec_results_tsp}
We expand the perspective of the proposed problem to taking \textit{route minimization} into consideration. We propose to characterize the route minimization as a travelling salesperson problem (TSP). Fig. \ref{fig_tsp_sa} demonstrates the EV chargers locations on the map of the State of Georgia. Through Figs. \ref{fig_tsp_sa_iter1} and \ref{fig_tsp_sa_iter300}, our SA algorithm (which has been shown as Algorithm \ref{alg_sa}) optimizes the route covering all the chargers and coming back to the starting point. The initial state in our algorithm is connection to a completely random neighbor. However, as the temperature is updated in each iteration, each node becomes able to connect to a closer neighbor. We notice here that our algorithm defines the temperature as the \textit{distance} between two stops on the map.

\begin{algorithm}
\fontsize{10}{10}\selectfont
\caption{SA implemented in this work}
\label{alg_sa}
$s = s_{0}$ \Comment*[r]{State initialization}
\For{$k \le k_{\text{max}}$}
{
    $\mathsf{T} \longleftarrow$ temperature$(\left(1 - \left(k+1\right)/k_{\text{max}}\right))$;\\
    
    $s_{\text{new}} \longleftarrow $ neighbor$\left(s\right)$;\\
    \If{$\mathsf{P}\left(\mathsf{E}\left(s\right),\mathsf{E}\left(s_{\text{new}}\right),\mathsf{T}\right) \ge \mathcal{U}\left(0,1\right)$}
    {
        $s \longleftarrow s_{\text{new}}$;
    }
}
\end{algorithm}

Algorithm \ref{alg_sa} presents the pseudocode for the SA implementation in this simulation. Notice of the parameters: $T$ for the temperature, $k$ for the index in the loop, $P\left(\cdot\right)$ for the acceptance probability, $E\left(\cdot\right)$ for the energy of a state; and $U$ for the uniform random variable. The name of the algorithm ``annealing'' comes from the metallurgy process, through which a metal cools and freezes into a crystalline structure with minimum energy \cite{sa_DuS02,sa_DuS02_MeR53}. SA starts with an initial solution at a higher temperature $T$, where the changes are accepted with higher probability $P$. So the exploration capability of the algorithm is high and the search space can be explored widely. As the algorithm continues to run, $T$ decreases gradually, like the annealing process, and the acceptance probability of a non-successful move $P$ decreases.

Fig. \ref{fig_tsp_distance} displays how the travelling distance converges as the iteration progresses via Algorithm \ref{alg_sa}.

\section{Conclusions and Future Work}\label{sec_conclusions}
This paper has formulated the optimal EV charger location problem into a capacitated FLP. Then, it laid out an effort to find an efficient method to solve the NP-hard problem. Via simulations, we presented the feasibility of using an ILP solver to solve the proposed problem. This paper also explored other numerical solving techniques: it presented a comparative quantification among several representative solving techniques. Lastly, in addition to the FLP, it also shed light on route minimization by solving a TSP. We found SA as an efficient technique to solve the TSP.

We plan to build on this work and progress on the front of \textit{societal impacts} of optimally located EV chargers. We have been developing a driving simulator \cite{arxiv2302_dhruba,arxiv2302_zach} that immerses users into a virtual reality-like interface and provides an actual experience of travelling through the optimally located EV chargers. This is expected to assist policy makers to communicate with general public in regard to deployment of EV chargers.


\end{document}